\documentclass[%
 reprint,
 amsmath,amssymb,
 aps,
]{revtex4-2}

\usepackage{graphicx}% Include figure files
\usepackage{dcolumn}% Align table columns on decimal point
\usepackage{bm}% bold math
\usepackage{xcolor}
\begin{document}

\preprint{APS/123-QED}

\title{Helicons in tilted-Weyl semimetals}% Force line breaks with \\

%\author{Panchlal Prabhat}
% \altaffiliation[Also at ]{Physics Department, L.N. Mithila University}%Lines break automatically or can be forced with \\
%\author{Amit Gupta}%
%\affiliation{%
 %Department of Physics, L.N. Mithila University, Darbhanga\\


\author{Shiv Kumar Ram}
\affiliation{ Department of Physics, Lalit Narayan Mithila University, Darbhanga, Bihar 846004, India}%

\author{Amit Gupta}
\affiliation{
  Department of Physics, M. R. M. College, Lalit Narayan Mithila University, Darbhanga, Bihar 846004, India}

%\affiliation{
 %Third institution, the second for Charlie Author
%}%


\date{\today}% It is always \today, today,

\begin{abstract}
We investigate the helicon modes in gapless tilted 3d Weyl semimetals(WSMs) within a semiclassical Boltzmann approach with the inclusion of orbital magnetic moment. These are transverse electromagnetic modes in three-dimensional (3D) electron systems in the presence of a static magnetic field. We calculate the all conductivity tensor and show that the helicon modes exist in tilted WSMs. The degeneracy of all the gapped collective modes at long wavelenth limit is unaffected by the tilting of the Weyl nodes.
\end{abstract} 

\maketitle

%\tableofcontents

\section{\label{sec:level1}
Introduction
%\protect\\ The line
 }
WSMs are topologically nontrivial conductors in which the valence and conduction bands touch at isolated points (the so-called Weyl nodes) in the Brillouin zone, locally forming Dirac cones \cite{wan2011topological, burkov2011weyl, xu2011chern, huang2015weyl, lv2015observation, xu2015discovery}. In general, Weyl nodes in solids do not behave exactly in the same manner as their high-energy Lorentz invariant analogs because they are tilted and anisotropic. Tilting the Weyl cones, meaning that the slope of the dispersion relation is not the same in opposite directions. Materials that exhibit such tilted Weyl cones are of type I if the tilt is relatively small and of type II if the cones are overtilted such that the electron and hole dispersions intersect the energy plane of the Weyl node itself. The tilt term influences the chiral anomaly, a phenomenon where chiral charge (related to the handedness of Weyl fermions) is not conserved in the presence of parallel electric and magnetic fields. This leads to observable effects like negative longitudinal magnetoresistance \cite{johansson2019chiral,
konye2021microscopic, ma2019planar,   das2019linear, mukherjee2018anomalous}.
The tilt modifies the optical conductivity, affecting the interband and intraband transitions. For example, the linear dependence of conductivity on photon energy, characteristic of Dirac and Weyl semimetals, is altered in the presence of tilt \cite{ishizuka2019tilting}. Tilting can also affect the anomalous Hall effect which arises from the Berry curvature of the electronic bands \cite{carbotte2016dirac, mukherjee2018anomalous, trescher2015quantum}, chiral magnetic effects \cite{van2017anisotropic}, the anomalous Nernst effect \cite{ferreiros2017anomalous, saha2018anomalous} and they modify the disorder effect \cite{sikkenk2019interplay}. The tilt can be tuned by various means such as structural distortions, spin textures or external fields allowing the manipulation of material properties \cite{konye2022horizon}. It is thus of considerable interest to investigate what such a tilt does to the helicon modes of a Weyl semimetal. \\

Helicons are low-frequency circularly polarized transverse electromagnetic waves propagating in a three-dimensional (3D) conducting medium under a static magnetic field 
($\bf B$) \cite{kittel2018introduction, jackson2021classical, konstantinov1960possible}. In conventional metals/plasmas, they follow a quadratic dispersion relation \cite{platzman1973waves}
\begin{equation}
\omega=\frac{k^2 B c}{4\pi n_e e}
\end{equation}

\noindent where $n_e$ is electron density and k is wave vector. These modes have been studied for topological gapless non-tilted WSM systems with the inclusion of orbital magnetic moment(OMM) \cite{pellegrino2015helicons}. The OMM can be thought as self-rotation of the Bloch wave packet and modify the energy of the Bloch electron under the external magnetic field \cite{sundaram1999wave}. Weyl semimetals introduce an axion term $\theta (\bf E \cdot \bf B $) in Maxwell’s equations due to their topological magnetoelectric effect, altering helicon dispersion \cite{pellegrino2015helicons}

\begin{equation}
\omega \sim \frac{k^2 B c}{4\pi n_e e} + k
\end{equation}
The second linear k-term originates due to the topological $\theta$ term. Helicon modes in Weyl semimetals are modified due to pseudofields (e.g., strain-induced gauge fields) which refer to pseudohelicons \cite{gorbar2017pseudomagnetic}.\\
 
In this paper, we investigate the effect of a finite tilt in WSMs to study helicon modes. Our main findings are that the optical conductivity tensor is non-trivially modified by a finite tilt and henceforth the dispersion of helicon modes is altered. Our results are valid for type-I WSMs. The organization of the paper is as follows: we start with semiclassical Boltzmann theory with the inclusion of magnetic orbital moment to evaluate the optical conductivity tensor in Sec.II. We evaluate the dispersion relation in Sec.III and finish with a discussion and conclusion in Sec.IV.\\

\section{Model Hamiltonian and Semiclassical Boltzmann approach}
The axion term $\theta(\textbf{r},t)=2(\textbf{b}\cdot\textbf{r}-b_{0}t)$, where $\bm{b}(b_0)$ denotes separation of nodes in momentum(energy) space introduce an additional term $\mathcal{L}_{\theta}$ in the standard Maxwell Lagrangian $\mathcal{L}$\cite{pellegrino2015helicons, zyuzin2012topological, deng2021exploring}

\begin{equation}
\mathcal{L} =\frac{1}{8\pi}(E^{2}-B^{2})-\rho \varphi+\textbf{J}\cdot\textbf{ A} + \mathcal{L}_{\theta}
\end{equation}
where
\begin{equation}
\mathcal{L}_{\theta}=-\frac{\alpha}{4\pi ^{2}} \theta(\textbf{r},t)\textbf{E}\cdot \textbf{B}
\end{equation}

\noindent where $\alpha= \frac{e^2}{\hbar c} \approx 1/137 $ is the usual fine-structure constant. The tilt term $\bm{t}$  will strikingly modify the inhomogeneous Maxwell equations due to the non-Minkowski space \cite{jalali2019electrodynamics}. The modification of the Maxwell's equations including both the tilt $\bm{t}$ and the axion terms  $\theta(\textbf{r},t)$ \cite{jalali2019electrodynamics}. 

\begin{equation}
\nabla \cdot \bm{D} = 4\pi (\rho + \frac{\alpha}{2\pi^{2}} \textbf{b }\cdot \textbf{B})+\textbf{t}\cdot\bf{ \nabla} \times \bf{B}
\end{equation}

\begin{equation}
\nabla \cdot {\bm B} = 0
\end{equation}

\begin{equation}
\nabla \times {\bm E}= -\frac{1}{c}\frac{\partial {\bm B}}{\partial t}
\end{equation}

\begin{eqnarray}
-&&\frac{\partial \bm{D}}{\partial t} +(1-t^2)\bm\nabla \times {\bm{ B}}-\frac{1}{c} \bm{t}\times \frac{\partial \bm{B}}{\partial t} +\bm{\nabla}\times (\bm{t}\times \bm{E})\nonumber\\-&&(\bm{t}\times \bm{\nabla})(\bm{t}\cdot \bm{B})
 =\frac{4\pi}{c}(\bm{J}-\frac{\alpha}{2\pi^2} \bm{b}\times \bm{E} +\frac{\alpha}{2\pi^{2}} b_{0} \bm{B})
\end{eqnarray}

We consider tilt $\bm t= (0,0,t_z)$ along the z-direction throughout this paper. From the above equations we can find the displacement field $\bm{D}$

\begin{eqnarray}\label{disp1.eqn}
\bm{D}=&&(\epsilon_\infty +\frac{4\pi i \sigma}{\omega}) \bm{E}+ \bm{t} \times \bm{B}-\frac{i c}{\omega} \bm{\nabla} \times (\bm{ t}\times \bm{E})+\nonumber\\&&
\frac{i c}{\omega} t^2 \bm{\nabla} \times \bm{B}+\frac{i c}{\omega} (\bm{t} \times \bm{\nabla})(\bm{t}\cdot \bm{ B})+\frac{ i\alpha }{\pi \omega}\dot{\theta}\bm{B}+\nonumber\\&&\frac{i c\alpha}{\pi \omega} (\bm{\nabla}\theta)\times \bm{E}
\end{eqnarray}

\noindent where $\epsilon_\infty$ is the static dielectric constant of the medium and $\sigma$ is the conductivity. The first term $(\epsilon_\infty +\frac{4\pi i \sigma}{\omega}) $ represents the plasmon oscillations of the bulk WSM. The magnetic field in Eq.(\ref{disp1.eqn}) is related to the electric field by Faraday's law of induction $\bm B=\frac{c}{i\omega}\nabla \times \bm E$. Accordingly, the wave equation for the electric field propagation is modified to be

\begin{eqnarray}
\bm{\nabla} \times (\bm{\nabla} \times \bm{E})+\frac{1}{c^{2}}\frac{\partial^2 }{\partial t^2}\bm{D}=0
\end{eqnarray}

In the vicinity of a nodal point with chirality s and Berry monopole charge of magnitude 1, the low-energy effective continuum Hamiltonian is given by \cite{dantas2018magnetotransport, nandy2021chiral, medel2024electric}
%%%%%%%%%%%%%%%%%
\begin{eqnarray} 
\mathcal{H} ( \textbf k) = \hbar v_F (s\bm{k}\cdot \bm{\sigma} +t_z k_z \sigma_0)
\end{eqnarray}

\noindent where $ \boldsymbol{\sigma} = \lbrace \sigma_x, \, \sigma_y, \, \sigma_z \rbrace $ the usual Pauli matrices and $v_F$ is the Fermi velocity.
The eigenvalues of the Hamiltonian are given by
\begin{align} 
\epsilon_{\mathbf k}^s
=\hbar v_F (t k_z + s k)
\end{align}
where $k=\sqrt{k_x^2+k_y^2+k_z^2}$ and the value $1$ ($-1$) for $s$ represents the conduction (valence) band. For a given chirality s of a single Weyl node, the semiclassical Boltzmann equation in equilibrium can be written as\\

\begin{eqnarray}
\frac{\partial \tilde{f}^s}{\partial t}+\bm{\dot{k}}.\frac{\partial \tilde{f}^s}{\partial \bm k}+\bm{\dot{r}}.\frac{\partial \tilde{f}^s}{\partial \bm r}=0\label{vlasov}
\end{eqnarray}

Here, $\tilde{f}^s$ is the electron distribution function. 

In the presence of a static magnetic field $\bf{B}$ and a time varying electric field $\bf{E}$, the semiclassical equations of motion are
\begin{eqnarray}
\bf{\dot{r}}=\frac{1}{\hbar}\bf{\nabla}_{\bf K}\tilde{\varepsilon}_{\bf k}^{s}-\bf{\dot{k}}\times{\Omega}_{k}^{s}\label{EoMa}\\
\hbar \bm{\dot{k}}=-e\bm{E}-\frac{e}{c}\dot{\bm{r}}\times \bm{B}\label{EoMb}
\end{eqnarray}

\noindent where -e is the electron charge, $\bf{E}$ and  $\bf{B}$ are external electric and magnetic fields, respectively. $\Omega_{k}^{s}$ is the Berry curvature. The first term on the right-hand side of Eq. (\ref{EoMa}) is $\bm{v}_{ \bm{k}}^s = \frac{1}{\hbar} \bm{\nabla}_k \tilde{\varepsilon}_{\bm{ k}}^s$, defined in terms of an effective band dispersion $\tilde{\varepsilon}_{s}(\bm{ k})$. In topological metals such as WSMs, this quantity acquires a term due to the intrinsic orbital moment
.i.e. $\tilde{\epsilon_{\bm k}^s}
=\epsilon_{\bm k}^s + \epsilon_{\bm k}^{m,s}$ with $\epsilon_{\bm k}^{m,s}  
= - \,{\bf B}_{s} \cdot \bf{m }_{s}  (\bf k)$. The $\bm{m}_{\bm{k}}^s$ is the orbital moment induced by the semiclassical “self-rotation” of the Bloch wavepacket. The velocity in k-space is defined as
\begin{eqnarray}
{\bf v}_{s}({\bf k} ) \equiv 
 \nabla_{{\bf k}} \tilde{\epsilon_{\bm k}^s}
 = {\bf  v}^{(0)} ({\bf k} ) + {\bf v}^{(m)}_s({\bf k} ) 
\end{eqnarray}

where ${\bf  v}^{(0)} ({\bf k} )= \frac{1}{\hbar} \bm{\nabla}_k \tilde{\varepsilon}^s$
and ${\bf  v}^{(m)}_s ({\bf k} )= \frac{1}{\hbar}\nabla_{{\bf k}} \epsilon_{\bm k}^{m,s}$.\\

 The $\Omega_{\bm{k}}^s$ is the Berry curvature and the orbital magnetic moment are defined as \cite{sundaram1999wave, xiao2010berry, gao2022suppression}

\begin{eqnarray}
\bm{\Omega}_{\bm{k}}^s&=&Im[\langle \bm{\nabla}_k u_k^s\vert \times \vert \bm{\nabla}_k u_k^s\rangle]\\
\bm{m}_{\bm{k}}^s&=&-\frac{e}{2\hbar}Im[\langle \bm{\nabla}_k u_k^s\vert \times (\mathcal{H}_J(\bm k)- \epsilon_{\bm k}^s)\vert \bm{\nabla}_k u_k^s\rangle] 
\end{eqnarray}
where $\vert u_k^s\rangle$ satisfies the equation $ \mathcal{H}(\bm k)\vert u_k^s\rangle= \epsilon_{\bm k}^s\vert u_k^s\rangle $.\\

The general expressions for Berry curvature and orbital magnetic moment for WSMs are \cite{nandy2021chiral}
\begin{eqnarray}
\bm{\Omega}_{\bm{k}}^s =\pm \frac{s}{2k^3}{\bm k}\\
\bm{m}_{\bm{k}}^s=\frac{s}{2} \frac{e v_F}{c k^2}{\bm k}
\end{eqnarray}

From these expressions, we immediately observe the identity
\begin{align}
\bm{m}_{\bm{k}}^s
=\frac{e v_F k}{c} \bm{\Omega}_{\bm{k}}^s 
\end{align}
%\textcolor{red}{
%While the BC changes sign with $s$, the OMM does not.}\\

By solving these coupled equations (\ref{EoMa}) and (\ref{EoMb}), one obtains
\begin{equation}
\dot{\bm{ r}}=\frac{1}{\hbar D}\Bigl[\nabla_{\bm{ k}}\tilde{\varepsilon}_{ \bm{k}}^s + e  \bm{E} \times \bm{\Omega}_{\bm k}^s + \frac{e}{c} (\nabla_\textbf{k}\tilde{\varepsilon}_{\bm{k}}^{s}\cdot\Omega_{\bm{k}}^{s} )\bm{B}\Bigr]
\label{eqforvel}
\end{equation}

\begin{equation}
\dot{ \bm{k}}=\frac{1}{\hbar D}[-e \bm{E}-\frac{e}{c} \nabla_{k}\varepsilon_{\bm k}^{s}\times \bm{B}-\frac{e^2}{\hbar c}(\bm{E} \cdot \bm{B})\bm{\Omega}_{\bm k}^{s}]
\end{equation}

\noindent where the factor $D=1+\frac{e}{\hbar c}( \bm{\Omega}_{\bm{k}}^s\cdot \bm{B})$ modifies the phase space volume.  For weak-magnetic-fields, we could Taylor expand $D^{-1}$ as
\begin{eqnarray}
D^{-1}=1+ \frac{e}{\hbar c}(\bm{\Omega}_{\bm k}^{s} \cdot \bm B) + \Bigl(\frac{e}{\hbar c}\Bigr)^2(\bm{\Omega}_{\bm k}^{s} \cdot \bm B)^2+....
\end{eqnarray}
Equation (\ref{vlasov}) can be solved by expanding the distribution function in a linear power in the electric field as follows:
\begin{equation}
\tilde{f}^{s}=\tilde{f}_{0}^{s}+\tilde{f}_{1}^{s} e^{-i\omega t} \label{distr}
\end{equation}

where $\tilde{f}_{1}^{s}$ is linear in $ \bf{E} $ and is parametrized as follows

\begin{equation}
\tilde{f}_{1}^{s}=-\frac{\partial\tilde f_{0}^{s}}{\partial \varepsilon_{k}^{s}}( X_{-}e^{i \phi}+ X_{+}e^{-i \phi} + X_{0}) \label{distr_exp}
\end{equation}

The $\tilde{f}_{0}^{s}( \epsilon_{\bm k}^s)$ can be expanded at low magnetic field as
\cite{pellegrino2015helicons}

\begin{eqnarray}
\tilde{f}_{0}^{s}(\tilde{\epsilon_{\bm k}^s})=\tilde{f}_{0}^{s}( \epsilon_{\bm k}^s -\bm{m}_{\bm{k}}^s\cdot \bm{B})\nonumber\\
\simeq \tilde{f}_{0}^{s}( \epsilon_{\bm k}^s)-\bm{m}_{\bm{k}}^s\cdot \bm{B}\frac{\partial \tilde{f}_{0}^{s}( \epsilon_{\bm k}^s)}{\partial \epsilon_{\bm k}^s}
\end{eqnarray}

From Eqs.(\ref{eqforvel}) and Eq.(\ref{distr_exp}), the expression for the current density at time t is given by
\begin{eqnarray}\label{cur_den}
\bm{j}_1=-\frac{e}{(2\pi)^3}\int d^3k \Bigl[\bm{\mathit{\tilde{v}}}_{\bm k}^s+\frac{e}{\hbar}(\Omega_{\bm k}^s \cdot \bm{\mathit{\tilde{v}}}_{\bm k}^s )\bm{B}\Bigr]\tilde{f}_1^{s} 
\end{eqnarray}

The above equation can be expressed in the frequency space $\omega$ as
\begin{equation}
j_a(\omega)=\sigma_{ab}(\omega)E_b(\omega)
\end{equation}
To include the effects from the OMM and the BC, we first define the quantities.

Inserting Eqs.(\ref{distr}) and (\ref{distr_exp}) in Eq.(\ref{vlasov}), we find

\begin{eqnarray}
\ X_{\pm}&=&\frac{e}{2D}\frac{{v_{k_\perp}^s}(E_{x}\pm iE_{y})}{ i[\omega\pm\frac{e B v_{k_\perp}^s}{ \hbar c D k_\perp}] }\\
&=& \frac{e}{2D}\frac{{v_{k_\perp}^s}(E_{x}\pm iE_{y})}{ i[\omega\pm\omega_{c}] }
\end{eqnarray}

where $ \omega_{c}=\frac{e Bv_{k_\perp}^s}{ \hbar c D k_\perp } $ is the effective cyclotron frequency in tilted WSM
and
\begin{equation}
X_{0}=\frac{e E_{z} }{ i \omega D }[\tilde{v_{kz}}+\frac{e B}{\hbar c}( \bm{\Omega}_{\bm{k}}^s\cdot \bm{\tilde{v}}_{\bm k}^s)]
\end{equation}

\noindent where $k_\perp=\sqrt{k_x^2+k_y^2}$  and $\omega_{c}=\frac{e B v_{k_\perp}^s}{\hbar c D k_\perp}$ is the general expressions for cyclotron frequeny of WSMs. 

Taking advantage of the azimuthal symmetry about the $z$-axis, we take advantage of the cylindrical coordinates defined by
\begin{align}
k_{x} = k_{\perp} \cos \varphi \,, \quad
k_{y} = k_{\perp} \sin \varphi\ \text{ and } k_{z} = k_{z}\,, 
\end{align}
where $k_{\perp} \in [0, \infty )$ and $\varphi \in [0, 2 \pi )$  \cite{medel2024electric}. We can rewrite the velocity components(x and y) as

\begin{equation}
 v_{k_x}^s=v_{k_\perp}^s \cos\varphi, v_{k_y}^s=v_{k_\perp}^s \sin\varphi
\end{equation}
with
\begin{eqnarray}
v_{k_\perp}^s&=&  \frac{v_F k_{\perp}}{ k}\Bigl[1-\frac{2s e B k_z}{ \hbar c k^3 }+\frac{e^2 B^2 k_z^2}{ \hbar^2 c^2 k^6 }\Bigr]^{1/2}\nonumber\\
\end{eqnarray}

In the next step, we change the variables from $(k_{\perp} , k_{z})$ to $(k, \theta )$ by the coordinate transformation \cite{dantas2018magnetotransport}.

\begin{eqnarray}
k_{\perp} = k \sin \theta \\
k_{z} = k \cos \theta 
\end{eqnarray}
The Jacobian of the transformation is $\mathcal{J} =  k $, leading to analytical expressions for longitudinal conductivities $\sigma_{zz}$ of tilted-WSMs

\begin{eqnarray}\label{Eq_cond_tilt}
\sigma_{zz}(\omega, t_z)
&=& \frac{i \mathcal{D}}{8 \pi t_z^3 \omega}\Bigl[-2t_z-\ln\Bigl(\frac{1-t_z}{1+t_z}\Bigr) +\frac{13 e^2B^2t_z^3 v_F^4\hbar^2}{30c^2\epsilon_F^4} \nonumber\\&+& \frac{e^2B^2t_z^5 v_F^4\hbar^2}{3c^2\epsilon_F^4}\Bigr]
\end{eqnarray}

\noindent where $\mathcal{D}=\frac{\pi e^2 n_e}{m_c}$ is the Drude weight, $n_e=\frac{\epsilon_F^3}{3\pi^2\hbar^3 v_F^3}$ is the electron density and $m_c=\frac{\epsilon_F}{v_F^2}$ represent WSM cyclotron mass. It is clearly seen from Eq.(\ref{Eq_cond_tilt}) that $\sigma_{zz}(\omega,t_z)$=$\sigma_{zz}(\omega,-t_z)$ i.e. independent of the sign of the tilt parameter. The above equation is also independent of s, therefore, the total conductivity has taken account of a factor of two.  Eq.(\ref{Eq_cond_tilt}) is the most important result of this section which reduces to the expression for the conductivity of the case without tilt in the limit $t_z\rightarrow 0$ \cite{pellegrino2015helicons}.

\begin{equation}
\sigma_{zz}(\omega)
=\frac{i \mathcal{D}}{2\pi\omega} \Bigl(1+\frac{13\hbar^2(\omega^0_c)^2}{20 \epsilon_F^2}
\Bigr)
\end{equation}

\begin{figure} [t] 
\includegraphics[scale=.5]{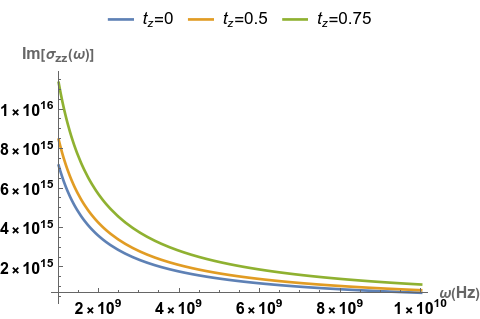} 
\caption{The frequency dependence of the longitudinal optical
conductivity at different tilt parameters. The other parameters are taken as $v_F =3\times 10^7 cm/s$, $\epsilon_F$ = 5 meV and B=.01 T.} 
\label{long_cond_freq}
\end{figure}

\begin{figure} [t] 
\includegraphics[scale=.5]{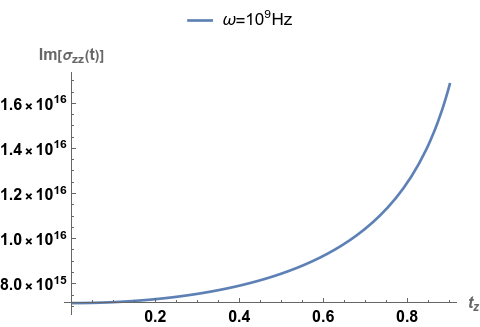} 
\caption{The tilt dependence of the longitudinal optical
conductivity at $\omega=10^{9} Hz$. The other parameters are taken as same in Fig.(\ref{long_cond_freq}).} 
\label{long_cond_tilt}
\end{figure}

The above component has been plotted as a function of frequency and tilt paramaters and frequency as shown in Figs.(\ref{long_cond_freq}) and (\ref{long_cond_tilt}) respectively for the parameters mentioned in caption of the figures. The longitudinal conductivity plot follow $\omega^{-1}$ relation. The conductivity increases monotonically with tilt parameter. The analytical expressions for transverse components of conductivities are not possible in low frequency limit.\\

Next, we calculate the transverse parts of conductivity for which we define cyclotron frequencies up to second order B are

\begin{eqnarray}
\omega_{c}& \approx &\frac{e B v_F^2(1+t_z \cos \theta)}{ \epsilon_F c(1-t_z \cos\theta)} -\frac{s e^2 B^2  \hbar v_F^4(\cos\theta -t_z)(1+t_z \cos \theta)^2}{2c^2 \sqrt{1-t^2}\epsilon_F^3(1-t_z \cos\theta)^4} \nonumber\\
&=&\omega_{c}^0 \frac{(1+t_z \cos \theta)}{(1-t_z \cos\theta)}+O(B^2)
\label{cyclo_freq_eff}
\end{eqnarray}

\noindent with $\omega_{c}^0=\frac{e B v_F^2}{ \epsilon_F c}$. The effective cyclotron frequency is modified due to tilt term. Eq.(\ref{cyclo_freq_eff}) is the second important result of this section which reduces to without tilt term in the limit $t\rightarrow0$ to
\cite{pellegrino2015helicons}
\begin{eqnarray}
\omega_{c} \approx \omega_{c}^0 -\frac{s e^2 B^2  \hbar v_F^4}{2c^2 \epsilon_F^3}\cos\theta
\label{cyclo_freq}
\end{eqnarray}

The leading power of B in Eq.(\ref{cyclo_freq_eff}) fixes the values of cyclotron frequencies $\omega_{c}^0=1.8\times10^{11}Hz$ for the parameter mentioned in the caption of Fig.(\ref{long_cond_freq}).

 \begin{figure} [t] 
\includegraphics[scale=.5]{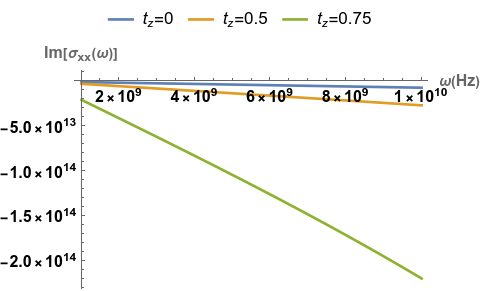} 
\caption{The frequency dependence of the transverse optical conductivity at different tilt parameters. The other parameters are taken as same in Fig.(\ref{long_cond_freq}).}
\label{trans_cond_freq}
\end{figure}

\begin{figure} [t] 
\includegraphics[scale=.5]{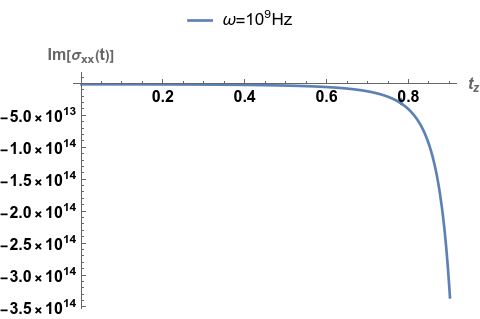} 
\caption{The tilt dependence of the transverse optical conductivity at $\omega=10^{9}Hz$. The other parameters are taken as same in Fig.(\ref{long_cond_freq}).}
\label{trans_cond_tilt}
\end{figure}

\begin{widetext}
\begin{eqnarray}
\sigma_{xx}(\omega, t_z)=\sigma_{yy}(\omega, t_z)=-\frac{e}{8\pi^3}\int\limits_{-\infty}^{\infty}\, dk_z
\int\limits_0^{\infty}k_\perp dk_\perp \int\limits_{0}^{\pi}\, d\varphi \frac{e}{2D} v_\perp^2 \cos \varphi^2\Bigl(\frac{1}{ i[\omega+\frac{e B v_\perp}{ \hbar cD k_\perp}] }+\frac{1}{ i[\omega-\frac{e B v_\perp}{\hbar c D k_\perp}] }\Bigr)\Bigl(-\frac{\partial f_0^s}{\partial \epsilon_{\bm k}^s}\Bigr) \\
\sigma_{xy}(\omega, t_z)=-\sigma_{yx}(\omega, t_z)=-\frac{e}{8\pi^3}\int\limits_{-\infty}^{\infty}\, dk_z
\int\limits_0^{\infty}k_\perp dk_\perp \int\limits_{0}^{\pi}\, d\varphi \frac{e}{2D} v_\perp^2 \sin \varphi^2\Bigl(\frac{1}{ i[\omega+\frac{e B v_\perp}{\hbar c D k_\perp}] }-\frac{1}{ i[\omega-\frac{e B v_\perp}{\hbar c D k_\perp}] }\Bigr)\Bigl(-\frac{\partial f_0^s}{\partial \epsilon_{\bm k}^s}\Bigr)
\end{eqnarray}
\end{widetext}

\begin{figure} [t] 
\includegraphics[scale=.5]{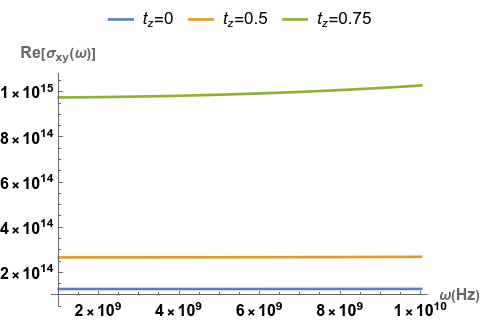} 
 \caption{The frequency dependence of the transverse Hall optical conductivity at different values of the tilt parameter. The other parameters are taken as same in Fig.(\ref{long_cond_freq}).} 
\label{Hall_xy_omega}
\end{figure}

\begin{figure} [t] 
\includegraphics[scale=.5]{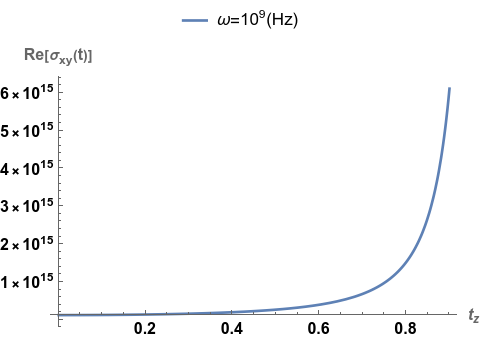} 
 \caption{The frequency dependence of the transverse Hall optical conductivity at $\omega=10^{9}Hz$. The other parameters are taken as same in Fig.(\ref{long_cond_freq}). } 
\label{Hall_xy_tilt}
\end{figure}

We solve the above equations numerically up to quadratic powers in B. Both the conductivity elements $\sigma_{xx}(\omega, t_z)$ and $\sigma_{xy}(\omega, t_z)$ are independent of sign of  the tilt parameter. Fig(\ref{trans_cond_freq}) and Fig.(\ref{Hall_xy_omega}) show their behavior as a function of $\omega$ while Fig(\ref{trans_cond_tilt}) and Fig.(\ref{Hall_xy_tilt}) show their behavior as a function of the tilt parameter.  We have plotted these figures in the low-frequency limit $\omega << \omega_c$ which is relevant for helicons. Both the xx and xy components have linear and flat dependnces with frequency respectively. However, their tilt dependence have similar behaviors and have asymtotically large values near $t\approx 1$. The frequnecy dependence of these conductivity elements increase with tilt parameter. These are key features of these components in this section. The remaining off-diagonal elements of the optical conductivity tensor vanish identically for symmetry reasons, independently of the frequency $\omega$.

\section{Helicons dispersion relation}
\begin{figure} [t] 
\includegraphics[scale=.5]{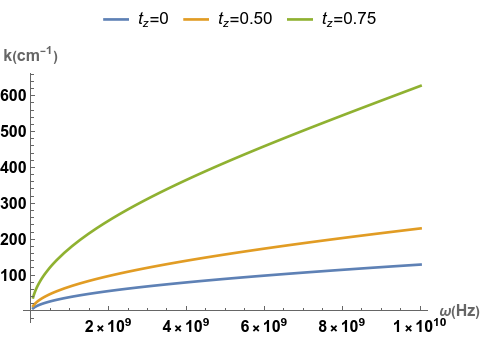} 
\caption{The $\omega$ vs. k dispersion at different values of parameters. The other parameters are taken as  $\epsilon_b=5$, $b_z=.01 \pi/(3.5 \times 10^{-8}) cm^{-1}$, $b_0=0$ and fine-structure constant $\alpha=1/137$. The other parameters are taken as same in Fig.(\ref{long_cond_freq}).} 
\label{dis_plot}
\end{figure}
Next, we derive the dielectric tensor from Eq.(\ref{disp1.eqn}) using relation $\bm D= \epsilon \bm E$
\begin{eqnarray}
\epsilon_{i j}&=& \delta_{i j}\epsilon_\infty + \frac{4\pi i}{\omega}\Bigl[\sigma_{i j}-\epsilon_{i j k}\frac{\alpha c}{2 \pi^2}\bigl(b_k-q_k\frac{b_0}{\omega}\bigr)\Bigr]\nonumber\\&+&
\Bigl(\frac{c}{\omega}\Bigr)^2 t^2\epsilon_{ipq}\epsilon_{qjl}k_pk_l -\Bigl(\frac{c}{\omega}\Bigr)^2 \epsilon_{ikl}\epsilon_{jpq}t_kt_pk_lk_q
\end{eqnarray}

\noindent where $\epsilon_{i j k}$ is the Levi-Civita antisymmetric tensor and the indices i, j and k run over the Cartesian coordinates x,y and z and $t^2=t_x^2+t_y^2+t_z^2$.  The above equation can be combined with the wave equation and gives the following relation

\begin{eqnarray}
(1-t_x^2-t_z^2)\Bigl(\frac{c k}{\omega}\Bigr)^2-\frac{2\alpha}{\pi c \omega}\Bigl(b_z-\frac{b_0 k}{\omega}\Bigr)&=&\epsilon_\infty \nonumber\\+\frac{4\pi i}{\omega} (\sigma_{xx}-i \sigma_{xy})
\label{dis_t}
\end{eqnarray}
From the above Eq.(\ref{dis_t}), the dispersion relations are even function of the tilt parameter(assuming $\bm t=t_z \bm \hat{z}$) and have been plotted in Fig.(\ref{dis_plot}). This plot shows the linear k dispersion(due to axion term) in addition to quadartic $k^2$ dispersion  which shows helicon modes exist with tilting the Weyl semimetals. With increasing tilt values, $\omega$ vs k relation increase in the same mannner(without affecting their power laws). This is our main result of this section.\\  

Next, we will look at the behavior of gapped collective modes at $\bm B=0$. 
The expressions for conductivities tilted along a general direction have been discussed in the literature \cite{detassis2017collective}

\begin{eqnarray}
\sigma_{xx}=\sigma_{yy}=\frac{e^2 \omega}{12\hbar v_F \pi}\Bigl[1-\theta(t_z-1)\Bigl(1-\frac{3t_z^2+1}{4t_z^3}\Bigr)\Bigr]
\end{eqnarray}
and
\begin{eqnarray}
\sigma_{zz}=\frac{e^2 \omega}{12\hbar v_F \pi}\Bigl[1-\theta(t_z-1)\Bigl(1-\frac{3t_z^2-1}{2t_z^3}\Bigr)\Bigr]
\end{eqnarray}

Note that the above expressions of the conductivities reduce the result of the untilted case for all values of $t_z<1$ (type-I WSMs)\cite{kargarian2015theory}. In the long-wavelength limit, we find that three gapped modes $\Omega_{\lambda}(q)$ with $\lambda=1,2,3$ are given by
\begin{eqnarray}
\Omega_{1}(k=0)=\omega_{-}\\
\Omega_{2}(k=0)=\omega_{p}/\sqrt{\epsilon_\infty}\\
\Omega_{3}(k=0)=\omega_{+}
\end{eqnarray}

\noindent where $\omega_{\pm}=\alpha c b/(\pi \epsilon_\infty) \pm \sqrt{
(\alpha c b)^2/(\pi \epsilon_\infty)^2+\omega_{p}^2/\epsilon_\infty}$ with $b=\mid\bf b\mid$ and $\omega_{p}$ define the plasma frequencies in WSMs
\begin{eqnarray}
\omega_{p}^2= \frac{4 e^2 (\omega_{c}^0)^2}{3 \pi \hbar v_F}
\end{eqnarray}

Therefore, the degeneracy of the three gapped collective modes at k=0 is lifted by the presence of the axion term but is unaffected by the tilt parameter in the electromagnetic response of the WSMs. \\

\section{Conclusion}
In summary, we have calculated the optical conductivity tensor of a 3D type-I tilted WSM by employing Boltzmann transport theory with the inclusion of orbital magnetic moment. The longitudinal and transverse parts of conductivity tensor have been obtained analytically and numerically respectively as a function of the tilt parameter. These elements increase with tilt parameter. We have demonstrated that helicon modes in type-I tilted WSMs with the inclusion of axion term in the standard Maxwell Lagrangian. These modes are a rich testbed for topological electrodynamics, blending condensed matter physics with electromagnetic wave theory. Their unique properties governed by chiral anomaly, Berry curvature and axion fields enable novel device concepts while posing intriguing theoretical challenges. Finally, we show that the degeneracy of the three gapped collective modes without magnetic field in the long wavelength limit is lifted by the presence of the axion term in the electromagnetic response of WSMs but is unaffected by a tilt parameter for $t<1$. It is very interesting to study the helicon modes for the type-II WSMs($t>1$) case.

%For deeper insights, consult studies on TaAs helicon experiments or axion electrodynamics in topological matter.

\section{Acknowledgements}
We thank Debanand Sa for fruitful discussions. The authors declared no financial support was received for this work.

\bibliography{helicon_tilted}
\end{document}